\documentstyle[12pt]{article}
\date{}
\def\be{\begin{equation}}
\def\ee{\end{equation}}
\def\bea{\begin{eqnarray}}
\def\eea{\end{eqnarray}}
\def\s{\sigma}

\def\de{\delta}
\def\om{\omega}
\def\pr{\prime}
\def\th{\theta}
\title{ Quasirotational disturbances\\
of linear string baryon configuration}
\author{G.\,S. Sharov\thanks{E-mail: german.sharov@tversu.ru}\\
{\small Tver state university}\\
{\small Tver, 170002, Sadovyj per. 35, Mathem. dep-t.}}
\begin{document}
\maketitle
\begin{abstract}
For the linear string baryon model $q$-$q$-$q$ the small disturbances
of its rotational motion (quasirotational states) are investigated.
The spectrum of these states is obtained in the form of Fourier series
and the complex eigenfrequencies are found in this spectrum.
So the classic rotational motions of the linear string baryon model
are unstable (unlike the similar motions for the string with massive ends).
This instability differs from its analog for the three-string baryon model.
\end{abstract}

\bigskip
\noindent{\bf Introduction}
\medskip

The string model of the meson is obvious from geometric point of view ---
this is the relativistic string with massive ends \cite{Ch}.
But for the baryon we have to choose between the following four string models
(four types of binding three quarks by relativistic strings)
suggested by X.~Artru in Ref.~\cite{AY}: a) the meson-like quark-diquark
model $q$-$qq$ \cite{Ko}; b) the ``three-string" model
or Y configuration with three strings from three quarks joined
in the fourth massless point \cite{CollinsPY}; c) the ``triangle" model
or $\Delta$-configuration with pairwise connection of three quarks
by three relativistic strings \cite{Tr};
d) the linear configuration $q$-$q$-$q$ with
quarks connected in series \cite{lin}

Here the latter model is considered (in comparison with some others).
It was not studied quantitatively before Ref.~\cite{lin} where
the initial-boundary value problem for classical
motion of this configuration were solved and the stability
problem for the rotational motion of this system was investigated.
This motion is the uniform rotation of the rectilinear string with the middle
quark at rest at a center of rotation \cite{Ko,4B}). Numerical experiments
in Ref.~\cite{lin} showed that the rotational motions of the system
$q$-$q$-$q$ are unstable. Any small asymmetric disturbances grow and result in
centrifugal moving away the middle material point (quark) and its complicated
motion with quasi-periodical varying of the distance between the nearest two
quarks. But the system $q$-$q$-$q$ is not transformed into the quark-diquark
($q$-$qq$) one, as was supposed previously in Ref.~\cite{Ko}.

In this paper for the system $q$-$q$-$q$ the result of the numerical
experiments in Ref.~\cite{lin} is proved analytically. For this purpose
the spectrum of quasirotational states (small disturbances of the rotational
motion) is obtained and compared with the similar spectrum for the string
with massive ends.

After the brief review the classical dynamics for the model $q$-$q$-$q$
in Sect.~1 we consider in Sect.~2 the quasirotational states of the
relativistic string with massive ends (the model $q$-$\overline q$ or
$q$-$qq$) and then in Sect.~3 the similar states of
the linear string baryon model $q$-$q$-$q$.

\bigskip
\noindent{\bf 1. Dynamics and rotational motions of the linear string
model $q$-$q$-$q$}
\medskip

Let's consider an open relativistic string with the tension $\gamma$
carrying three pointlike masses $m_1$, $m_2$, $m_3$ (the masses $m_1$
and $m_3$ are at the ends of the string).
The action for this system is \cite{lin}
\be
 S=-\int\limits_{\tau_1}^{\tau_2}\! d\tau\left\{\gamma\!
\int\limits_{\s_1(\tau)}^{\s_3(\tau)}\!\!
\left[\big(\dot X,X'\big)^2-\dot X^2X'{}^2\right]^{1/2}\!d\s+\sum
_{i=1}^3m_i\sqrt{\dot x_i^2(\tau)}\right\}.
\label{S}\ee
Here $X^\mu(\tau,\s)$ are coordinates of a point of
the string in $D$-dimensional Minkowski space $R^{1,D-1}$
with signature $(+,-,-,\dots)$,
the speed of light $c=1$,
$\,(\tau,\s)\in G= G_1\cup G_2$ (Fig.\,1),
$\big( a,b\big)=a^\mu b_\mu$ is the (pseudo)scalar product,
$\dot X^\mu=\partial_\tau X^\mu$, $X'{}^\mu=\partial_\s X^\mu$,
$\dot x_i^\mu(\tau)=\frac d{d\tau}X^\mu(\tau,\s_i(\tau))$; $\s_i(\tau)$
($i=1,2,3$) are inner coordinates of world lines for
quarks\footnote{We use the term
``quark" for brevity, here and below quarks, antiquarks and diquarks
are material points on the classic level.}
shown in Fig.~1. If $m_2=0$ the action (\ref{S}) will describe
the relativistic string with massive ends \cite{Ch}.

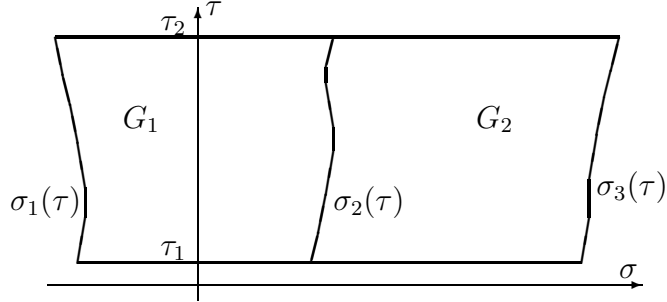
\begin{figure}[th]
\unitlength=1.0mm
\begin{center}
\begin{picture}(86,40)
\put(5,2){\vector(1,0){79}} \put(25,0){\vector(0,1){39}}
\thicklines
\put(9,5){\line(1,0){67}} \put(6,35){\line(1,0){75}}
\put(9,5){\line(1,6){1}} \put(10,11){\line(0,1){4}}
\put(10,15){\line(-1,6){1}} \put(9,21){\line(-1,5){1}}
\put(8,26){\line(-1,4){1}} \put(7,30){\line(-1,5){1}}
\put(40,5){\line(1,4){1}} \put(41,9){\line(1,5){1}}
\put(42,14){\line(1,6){1}} \put(43,20){\line(0,1){3}}
\put(43,23){\line(-1,6){1}} \put(42,29){\line(0,1){2}}
\put(42,31){\line(1,4){1}}
\put(76,5){\line(1,6){1}} \put(77,11){\line(0,1){5}}
\put(77,16){\line(1,6){1}} \put(78,22){\line(1,5){1}}
\put(79,27){\line(1,4){2}}
\put(81,3){$\sigma$} \put(26,38){$\tau$}
\put(20,36){$\tau_2$} \put(20,6){$\tau_1$}
\put(0,12){$\sigma_1(\tau)$} \put(43,12){$\sigma_2(\tau)$}
\put(78,14){$\sigma_3(\tau)$} \put(15,23){$G_1$}
\put(62,23){$G_2$}
\end{picture}
\caption{Domain of integration in Eq.~(1).}
\end{center}
\end{figure}

The equations of motion of the $q$-$q$-$q$ string and the boundary conditions
are derived from the action (\ref{S}) \cite{lin}. They take
the simplest form if with the help of nondegenerate reparametrization
$\tau=\tau(\tilde\tau,\tilde\s)$, $\s=\s(\tilde\tau,\tilde\s)$
the induced metric on the world surface of the string is made
continuous and conformally-flat \cite{Tr}, i.e., satisfies the orthonormality
conditions
\be\dot X^2+X'{}^2=0,\qquad\big(\dot X,X'\big) = 0.
\label{ort}\ee

Under conditions (\ref{ort}) the equations of motion become linear
\be \ddot X^\mu-X''{}^\mu=0
\label{eq}\ee
and the boundary conditions take the simplest form
\bea
&\displaystyle m_i\frac d{d\tau}U^\mu_i(\tau)+\epsilon_i
\gamma\bigl[X'{}^\mu{}+\s_i'(\tau)\,\dot X^\mu\bigr]\bigg|_{\s=\s_i(\tau)}=0,
\quad i=1,3,&\label{qq}\\
&\displaystyle m_2\frac d{d\tau}U^\mu_2(\tau)-\gamma\big[X'{}^\mu+
\s_2'(\tau)\,\dot X^\mu\big]\bigg|_{\s=\s_2(\tau)+0}+
\gamma\big(X'{}^\mu+\s_2'(\tau)\,\dot X^\mu\big)
\bigg|_{\s=\s_2(\tau)-0}=0.&\label{qqq}
\eea
Here $\epsilon_1=-1$, $\epsilon_3=1$ and
\be
U^\mu_i(\tau)=\frac{\dot x_i^\mu(\tau)}{\sqrt{\dot x_i^2(\tau)}}=
\frac{\dot X^\mu+\s_i'(\tau)\,X'{}^\mu}
{\sqrt{\dot X^2\cdot(1-\s_i'{}^2)}}\bigg|_{\s=\s_i(\tau)},\quad i=1,\,2,\,3
\label{Ui}\ee
are the unit $R^{1,D-1}$-velocity vector of $i$-th quark.

Derivatives of $X^\mu(\tau,\s)$ can have discontinuities
on the line $\s=\s_2(\tau)$.
However, the function $X^\mu(\tau,\s)$ and the tangential derivatives
$\frac d{d\tau}X^\mu(\tau,\s_2(\tau))$ are continuous. In Ref.~\cite{lin}
we showed that without loss of generality one can choose the coordinates
$\tau,\,\s$ satisfying both the orthonormality conditions (\ref{ort})
and the following restrictions for the endpoints' inner equations:
\be
\s_1(\tau)=0,\qquad \s_3(\tau)=\pi\;\;\;\Longrightarrow\;\;\;\s\in[0,\pi].
\label{s0pi}\ee
But we can't fix the remaining function $\s=\s_2(\tau)$ for the middle quark
in general if the conditions (\ref{ort}) and (\ref{s0pi}) are satisfied.

However, the rotational motion of this configuration (uniform rotation of
the rectilinear string segment with the middle quark at the rotational center)
is the well known exact solution of equations (\ref{eq}) satisfying all
conditions (\ref{ort}), (\ref{qq})\,--\,(\ref{s0pi}). This solution may be
presented in the form \cite{stabPRD}
\be
X^\mu=X_{rot}^\mu(\tau,\s)=\Omega^{-1}
\big[\theta\tau e_0^\mu+\cos(\theta\s+\phi_1)\cdot
e^\mu(\tau)\big],\qquad\s\in[0,\pi],
\label{rot}\ee
Here $\Omega$ is the angular velocity,
$e_0^\mu$ is the unit time-like velocity vector of c.m. in Minkowski
space, $e^\mu(\tau)=e_1^\mu\cos\theta\tau+e_2^\mu\sin\theta\tau$ is the unit
($e^2=-1$) space-like rotating vector directed along the string,
The parameters $\theta$ (dimensionless frequency) and $\phi_1$ are connected
with the constant speeds $v_i$ of the ends
\be
v_1=\cos\phi_1,\qquad v_3=-\cos(\pi\theta+\phi_1),\qquad
m_i\Omega/\gamma=v_i^{-1}-v_i,\quad i=1,3.
\label{RvOm}\ee
The central massive point of the $q$-$q$-$q$ system is at rest (in the
corresponding frame of reference) at the rotational center.
Its inner coordinate is
\be
\s_2(\tau)=\s_2^{rot}=(\pi/2-\phi_1)/\theta={\mbox{const}}.
\label{sig2}\ee

In the following two chapters we'll study small disturbances
of the rotational motion (\ref{rot}) (quasirotational states). But before
the analysis of the complicated linear string baryon model $q$-$q$-$q$
in Sect.~3  we'll consider in the following section
the more simple system --- the relativistic string with massive ends
$q$-$\overline q$.

\bigskip
\noindent{\bf 2. Quasirotational motions of the string with massive ends}
\medskip

The quasirotational states of various string hadron models \cite{4B,stabPRD}
are interesting due to the following reasons:
(a) we are to search the motions describing the hadron states, which are
usually interpreted as higher radially excited states and other states
in the potential models, in other words, we are to describe
the daughter Regge trajectories;
(b) the quasirotational states are the basis for quantization of these
nonlinear problems in the linear vicinity of the solutions (\ref{rot}),
(if they are stable);
(c) the quasirotational motions are necessary for solving the important
problem of stability of rotational states for all string hadron models.

For the meson string model the quasirotational motions of slightly curved
string with massive ends were studied in Refs.~\cite{AllenOV}. But these
authors used very narrow ansatz for searching these disturbances
and the complicated nonlinear form of the string motion equations beyond
the conditions (\ref{ort}). Besides they neglected some important dependencies
and the boundary conditions (\ref{qq}) so these solutions in
Refs.~\cite{AllenOV} were not correct (details are in Ref.~\cite{stabPRD}).

In Ref.~\cite{stabPRD} another approach for obtaining
the quasirotational solutions was suggested. It includes the
orthonormality conditions (\ref{ort}) and, hence,
the linear equations of motion (\ref{eq})
with their general solution
\be
X^\mu(\tau,\s)=\frac1{2}\big[\Psi^\mu_+(\tau+\s)+\Psi^\mu_-(\tau-\s)\big].
\label{gen}\ee
So the problem is reduced to the system of ordinary differential equations
resulting from the boundary conditions (\ref{qq}).
The unknown function may be $\Psi^\mu_+(\tau)$, $\Psi^\mu_-(\tau)$, or
unit velocity vectors of the endpoints $U^\mu_1(\tau)$ or $U^\mu_2(\tau)$
of the string with massive ends
--- this is equivalent due to the relations \cite{PeSh}
\be
\Psi^{\pr\mu}_\pm(\tau\pm\s_i)=m_i\gamma^{-1}\big[
\sqrt{-U_i^{\pr2}(\tau)}\,U_i^\mu(\tau)\mp(-1)^i U_i^{\pr\mu}(\tau)\big].
\label{psdet}\ee

The expression (\ref{rot}) describes the rotational motion not only for
the baryon $q$-$q$-$q$ model but also for the relativistic string with
massive ends (where $\s_2=\pi$ and $v_3$ in Eq.~(\ref{RvOm}) should be
substituted by $v_2$). In Ref.~\cite{stabPRD} the boundary conditions
(\ref{qq}) for this model with one infinitely heavy (fixed) end with
$m_2\to\infty$ on the basis of relations (\ref{gen}), (\ref{psdet})
were reduced to the ordinary differential equation with respect to the
vector function $U_1^\mu(\tau)$.

For the case with two non-zero finite masses ($0<m_i<\infty$)
the generalization of this equation resulting from Eqs.~(\ref{qq}),
(\ref{psdet}) takes the form
\be
\begin{array}{c}
U^{\pr\mu}_1(\tau)=m_2m_1^{-1}\big[\de^\mu_\nu-
U_1^\mu(\tau)\,U_{1\nu}(\tau)\big]\Big[\sqrt{-U_2^{\pr2}(\tau-\pi)}\,
U_2^\nu(\tau-\pi)-U_2^{\pr\nu}(\tau-\pi)\Big],\\
U^{\pr\mu}_2(\tau)=m_1m_2^{-1}\big[\de^\mu_\nu-
U_2^\mu(\tau)\,U_{2\nu}(\tau)\big] \Big[\sqrt{-U_1^{\pr2}(\tau-\pi)}\,
U_1^\nu(\tau-\pi)-U_1^{\pr\nu}(\tau-\pi)\Big],
\rule[4.3mm]{0mm}{1mm}\end{array}
\label{sysU}\ee
where $\de^\mu_\nu=\left\{\begin{array}{ll}1,&\mu=\nu,\\0,&\mu\ne\nu.
\end{array}\right.$
This system of ordinary differential equations with shifted arguments
exhaustively describes the classic dynamics of the string with massive ends.

If the vector-function $U_1^\mu(\tau)$ (or $U_2^\mu(\tau)$) is given in the
segment $I=[\tau_0,\tau_0+2\pi]$ (the values $\gamma/m_i$,
$U_2^\mu(\tau_0+\pi)$ are also given)
one can determine the functions $U_i^\mu(\tau)$ for $\tau>\tau_0$ from the
system (\ref{sysU}). Then we may obtain the world surface $X^\mu(\tau,\s)$ with the
help of the relations (\ref{psdet}) and (\ref{gen}). So we may conclude that
the function $U_1^\mu(\tau)$ or $U_2^\mu(\tau)$ given in the segment
$I$ contains all information about this motion of the system \cite{stabPRD}.

For the rotational motion (\ref{rot}) the velocities $U_i^\mu$ of the moving quark
satisfying Eqs.~(\ref{sysU}) may be written in the form
\be\!\!
U_1^\mu= U_{1(rot)}^\mu(\tau)=\Gamma_1\big[e_0^\mu+v_1\acute e^\mu(\tau)\big],
\quad U_2^\mu= U_{2(rot)}^\mu=\Gamma_2\big[e_0^\mu-v_2\acute e^\mu(\tau)\big],
\quad\Gamma_i=(1-v_i^2)^{-1/2}\!.\!\!\!
\label{Urot}\ee
Here the unit space-like rotating vectors $\acute e^\mu$ and $e^\mu$
\be
e^\mu(\tau)=e_1^\mu\cos\theta\tau+e_2^\mu\sin\theta\tau,
\qquad\acute e^\mu=\theta^{-1}\frac d{d\tau}e^\mu(\tau)=
-e_1^\mu\sin\theta\tau+e_2^\mu\cos\theta\tau
\label{evec}\ee
consist the moving basis in the rotational plane. The four vectors $e_0^\mu$,
$e^\mu(\tau)$, $\acute e^\mu(\tau)$, $e_3^\mu$ will be used below as the
orthonormal tetrade in the Minkowski space $R^{1,3}$.

To study the small disturbances of the rotational motion (\ref{rot})
we consider arbitrary small disturbances of this motion or of the vector
(\ref{Urot}) in the form
\be
U^\mu_i(\tau)=U^\mu_{i(rot)}(\tau)+u_i^\mu(\tau),\qquad |u_i^\mu|\ll1.
\label{U+u}\ee
For the exhaustive description of this quasirotational state the disturbance
$u_i^\mu(\tau)$ may be given in the initial segment $I=[\tau_0,\tau_0+2\pi]$.
It is small so we neglect in the linear approximation the second order terms.
The equality $U_i^2(\tau)=1$ for both vectors $U_i^\mu$ and $U_{i(rot)}^\mu$
leads in the linear approximation to the condition
\be
U_{i(rot)}^\mu(\tau)\,u_{i\mu}(\tau)=0.
\label{Uu}\ee

When we substitute the expressions (\ref{U+u}) into the system (\ref{sysU})
and omit the second order terms we obtain the linearized system of equations
describing the evolution of small arbitrary disturbances $u_i^\mu$.
Considering projections of these two vector equations onto the basic vectors
$e_0$, $e$, $\acute e$, $e_3$, we obtain the following system of equations
with respect to projections of $u_i^\mu$:
\be
\!\!\!\!\!\!\begin{array}{c}
u'_{10}(\tau)+Q_1u_{10}(\tau)-\Gamma_1Q_1u_{1e}(\tau)=
M_0\big[u'_{20}-Q_2u_{20}+\Gamma_2Q_2u_{2e}\big],\\
\!\!u'_{1e}(\tau)+Q_1u_{1e}(\tau)+\theta v_1^{-1}u_{10}(\tau)=
M_1^{-1}\big[-u'_{2e}-Q_1u_{2e}+N_2^*u'_{20}+N_2u_{20}\big],
\rule[3.3mm]{0mm}{1mm}\!\!\\
u'_{20}+Q_2u_{20}+\Gamma_2Q_2u_{2e}=
M_0^{-1}\big[u'_{10}(-)-Q_1u_{10}(-)-\Gamma_1Q_1u_{1e}(-)\big],
\rule[3.3mm]{0mm}{1mm}\\
\!\!\!\!u'_{2e}+Q_2u_{2e}-\theta v_2^{-1}u_{20}=
M_1\big[-u'_{1e}(-)-Q_2u_{1e}(-)+N_1^*u'_{10}(-)+N_1u_{10}(-)\big],
\rule[3.3mm]{0mm}{1mm}\!\!\\
u'_{1z}(\tau)+Q_1u_{1z}(\tau)=(m_2/m_1)\big[-u'_{2z}+Q_2u_{2z}\big],
\rule[3.3mm]{0mm}{1mm}\\
u'_{2z}+Q_2u_{2z}=(m_1/m_2)\big[-u'_{1z}(-)+Q_1u_{1z}(-)\big].
\rule[3.3mm]{0mm}{1mm}\end{array}\!\!\!
\label{sysu}\ee
Here $Q_i=\Gamma_i\theta v_i={}$const, $(-)\equiv(\tau-2\pi)$, the functions
\be
u_{i0}(\tau)=(e_0,u_i),\qquad u_{ie}(\tau)=(e,u_i),\qquad
u_{iz}(\tau)=(e_3,u_i)
\label{u0ez}\ee
are the projections of the vectors $u_i^\mu(\tau)$ onto the mentioned basis.
The projections of $u_i^\mu$ onto $\acute e^\mu$ may be expressed through
$u_{i0}$: $(\acute e,u_i)=(-1)^iv_i^{-1}u_{i0}$ due to the equality
(\ref{Uu}). Arguments $(\tau-\pi)$ of the functions $u_{20}$, $u_{2e}$,
$u_{2z}$ are omitted.
The constants in Eqs.~(\ref{sysu}) are
$$\begin{array}{c}
M_0=m_2Q_1/(m_1Q_2),\qquad M_1=m_1\Gamma_1/(m_2\Gamma_2),\\
N_i^*=-(-1)^i(1+Q_{3-i}/Q_i)/\Gamma_i,\qquad
N_i=(-1)^i(Q_{3-i}+Q_i\kappa_i)/\Gamma_i,\qquad\kappa_i=1+v_i^{-2}.
\rule[3.3mm]{0mm}{1mm}\end{array}$$

We shall search solutions of the linearized system (\ref{sysu}) in the form
\be
u_i^\mu=A_i^\mu e^{-i\om\tau}.
\label{uharm}\ee
For the last two equations (\ref{sysu}) (they form the closed subsystem)
solutions in the form (\ref{uharm}) exist only if the dimensionless
frequency $\om$ satisfies the transcendental equation
\be
\frac{\om^2-Q_1Q_2}{(Q_1+Q_2)\,\omega}=
\cot\pi\omega,
\label{zfreq}\ee
but for the subsystem of the first 4 equations (\ref{sysu})
the corresponding frequencies $\om=\tilde\om$ are roots of another equation
\be
\frac{(\tilde\om^2-Q_1^2\kappa_1)(\tilde\om^2-Q_2^2\kappa_2)
-4Q_1Q_2\tilde\om^2}
{2\tilde\om\big[Q_1(\tilde\om^2-Q_2^2\kappa_2)+
Q_2(\tilde\om^2-Q_1^2\kappa_1)\big]}=\cot\pi\tilde\om,
\label{pfreq}\ee
One can numerate the roots $\om=\om_n$ of Eq.~(\ref{zfreq}) and
$\tilde\om=\tilde\om_n$ for Eq.~(\ref{pfreq}) in order of increasing so that
$\om_0=\tilde\om_0=0$, $n-1<\om_n,\tilde\om_n<n$ for $n\ge1$.
The system of functions $\exp(-i\om_n\tau)$ or $\exp(-i\tilde\om_n\tau)$,
$n=0,\,\pm1,\,\pm2,\dots$ is the full system in the class $C(I)$
\cite{stabPRD,PeSh} so an arbitrary continuous functions $u(\tau)$
in this segment with the length
$2\pi$ may be expanded in the Fourier series
\be
u(\tau)=\sum_{n=-\infty}^{+\infty}u_n\exp(-i\om_n\tau),\qquad
\tau\in I=[\tau_0,\tau_0+2\pi].
\label{uFour}\ee
Using this expansion for the disturbance (\ref{U+u}) $u_i^\mu$ of the
velocity vectors (\ref{Urot}) we obtain with the help of
Eqs.~(\ref{gen}), (\ref{psdet}) the following expression for an arbitrary
quasirotational motion of the string with massive ends \cite{Exc}:
\begin{eqnarray}
&\displaystyle
X^\mu(\tau,\s)=X^\mu_{rot}(\tau,\s)+\!\sum_{n=-\infty}^\infty
\Big\{e_3^\mu\alpha_n\cos(\om_n\s+\phi_n)\exp(-i\om_n\tau)&
\nonumber\\
&\qquad\qquad{}+\beta_n\big[e_0^\mu f_0(\s)+e_\perp^\mu(\tau) f_\perp(\s)+
ie^\mu(\tau) f_r(\s)\big]\exp(-i\tilde\om_n\tau)\Big\}.&
\label{osc}
\end{eqnarray}
Each term in Eq.~(\ref{osc}) describes the
string oscillation that looks like the stationary wave with $n$ nodes.
There are two types of these stationary waves: (a) orthogonal oscillations
along $z$ or $e_3$-axis at the frequencies proportional to the roots
$\om_n$ of equation\footnote{It is interesting that the same equation
(\ref{zfreq}) describes the spectrum
of states for the relativistic string with massive ends with linearizable
boundary conditions \cite{PeSh}.} (\ref{zfreq}),
and (b) planar oscillations (in the rotational plane $e_1,e_2$)
at the dimensionless frequencies $\tilde\om_n$ satisfying the equation
(\ref{pfreq}) with the following expressions for $f_0$, $f_\perp$, $f_r$:
\be
\begin{array}{c}
f_0(\s)=\frac12(Q_1\kappa_1\tilde\om_n^{-1}-Q_1^{-1}\tilde\om_n)
\cos\tilde\om_n\s-\sin\tilde\om_n\s,\\
f_\perp(\s)=\Gamma_1(\Theta_n\tilde\om_n-h_nv_1)\,C_\th C_\om
-v_1^{-1}C_\th S_\om-\Gamma_1\th\Theta_n S_\th S_\om+h_n S_\th C_\om,
\rule[3.5mm]{0mm}{1mm}\\
f_r(\s)=\Gamma_1(\Theta_n\tilde\om_n-h_nv_1)\,S_\th S_\om
+v_1^{-1}S_\th C_\om+\Gamma_1\th\Theta_n C_\th C_\om-h_n C_\th S_\om.
\rule[3.5mm]{0mm}{1mm}\end{array}
\label{f0pr}\ee
Here $\displaystyle
\Theta_n=\frac{2\th}{\tilde\om_n^2-\th^2},\;\;\;
h_n=\frac12\Big[\frac\th{\tilde\om_n}\Big(\frac1{v_1}+v_1\Big)+
\frac{\tilde\om_n}\th\Big(\frac1{v_1}-v_1\Big)\Big]\;$,
$C_\th(\s)=\cos\th\s$, $S_\th(\s)=\sin\th\s$, $C_\om(\s)=\cos\tilde\om_n\s$,
$S_\om(\s)=\sin\tilde\om_n\s$.

The frequencies $\om_n$ and $\tilde\om_n$ from Eqs.~(\ref{zfreq}) and
(\ref{pfreq}) are real numbers
so the rotations (\ref{rot}) of the string with massive ends are stable
in the linear approximation. One may consider the expansion (\ref{osc})
for an arbitrary quasirotational motion as the basis for further quantization
of this system in the linear vicinity of the solution (\ref{rot}).

\bigskip
\noindent{\bf 3. Quasirotational motions of the linear string
model $q$-$q$-$q$}
\medskip

The stability problem for the rotational motion (\ref{rot}) of the linear
system $q$-$q$-$q$ was studied numerically in Ref.~\cite{lin}.
Now we present the analytical investigation of this problem based
upon the method developed in the previous section for the string
with massive ends. In particular, we may express the world surface of
the linear $q$-$q$-$q$ configuration through the unit velocity vectors
(\ref{Ui}) $U_1^\mu(\tau)$ and $U_3^\mu(\tau)$ of the massive end
using the generalized formula (\ref{psdet})
\be
\begin{array}{c}
\Psi^{\pr\mu}_{1\pm}(\tau\pm\s_2)=m_1\gamma^{-1}\Big[
\sqrt{-U_1^{\pr2}(\tau\pm\s_2)}\,U_1^\mu(\tau\pm\s_2)\pm
U_1^{\pr\mu}(\tau\pm\s_2)\big],\\
\Psi^{\pr\mu}_{2\pm}(\tau\pm\s_2)=m_3\gamma^{-1}\Big[
\sqrt{-U_3^{\pr2}(\tau\pm\s_2\mp\pi)}\,U_3^\mu(\tau\pm\s_2\mp\pi)
\mp U_3^{\pr\mu}(\tau\pm\s_2\mp\pi)\big].
\rule[4.3mm]{0mm}{1mm} \end{array}
\label{pslin}\ee
Here
\be
X^\mu(\tau,\s)=\frac1{2}\big[\Psi_{i+}^\mu(\tau+\s)+\Psi_{i-}^\mu(\tau-\s)
\big],\qquad (\tau,\s)\in G_i,
\label{genlin}\ee
is the general solution of the string equation (\ref{eq})
(generalization of Eq.~(\ref{gen})) for this system. It is described by
the different functions $\Psi_{1\pm}^\mu$ and $\Psi_{2\pm}^\mu$ in the domains
$G_1$ and $G_2$ in Fig.~1, because $X^\mu(\tau,\s)$ is not continuous
on the line $\s=\s_2(\tau)$, dividing these domains.
However, as was mentioned above the function $X^\mu(\tau,\s)$ and the
tangential derivatives
$\frac d{d\tau}X^\mu(\tau,\s_2(\tau))$ are continuous.
The latter fact results in the equality
\be
(1+\s'_2)\,\Psi^{\pr\mu}_{1+}(+)+(1-\s'_2)\,\Psi^{\pr\mu}_{1-}(-)=
(1+\s'_2)\,\Psi^{\pr\mu}_{2+}(+)+(1-\s'_2)\,\Psi^{\pr\mu}_{2-}(-).
\label{cont15}\ee
Here $(+)\equiv\big(\tau+\s_2(\tau)\big)$,
$(-)\equiv\big(\tau-\s_2(\tau)\big)$, the conditions (\ref{s0pi}) are assumed.

If we substitute the general solution (\ref{genlin}) into the boundary
condition of the middle quark (\ref{qqq}) it will take the form
\be
m_2\frac d{d\tau}U^\mu_2(\tau)=\gamma\big[\de^\mu_\nu-U_2^\mu(\tau)\,
U_{2\nu}(\tau)\big]\big[(1+\s_2')\,\Psi_{2+}^{\pr\nu}(\tau+\s_2)+
(1-\s_2')\,\Psi_{1-}^{\pr\nu}(\tau-\s_2)\big].
\label{qqqli}\ee

The analog of the system (\ref{sysU}) for the model $q$-$q$-$q$ may be
obtained if we substitute the expressions (\ref{pslin}) into the boundary
conditions (\ref{cont15}) and (\ref{qqqli}). This system of equations
(it is too cumbrous so is isn't written here explicitly)
(\ref{qqqli}), (\ref{cont15}), (\ref{pslin}) connects the fuctions
$U_i^\mu(\tau)$, $i=1,2,3$ and $\s_2(\tau)$.
For analysis of the quasirotational states and the stability problem
for the motion (\ref{rot}) of the $q$-$q$-$q$ model we substitute
into the mentioned system of equations the small disturbances of the
velocity vectors $U_i^\mu$ in the same form (\ref{U+u}), $i=1,2,3$
omitting the 2-nd order terms of $u_i^\mu(\tau)$. Besides (this is specific
feature of the model $q$-$q$-$q$) one should consider the small disturbance
of the function $\s_2(\tau)$ in this system
$$
\s_2(\tau)=\s_2^{rot}+\de\s_2(\tau).$$
Here for the rotational motion (\ref{rot}) $\s_2^{rot}$ is the value
(\ref{sig2}) and the velocities $U_{i(rot)}^\mu$ are described by the
slightly modified formula (\ref{Urot})
$$
U_{i(rot)}^\mu(\tau)=\Gamma_i\big[e_0^\mu-\epsilon_iv_i\acute
e^\mu(\tau)\big],\qquad\epsilon_1=-1,\quad\epsilon_3=1,
$$
where $v_2=0$, because the middle quark with the mass $m_2$
is at rest at the rotational center.

Searching oscillatory solutions of this linearized system
(the analog of Eqs.~(\ref{sysu})), we substitute into it the small
disturbances $\de\s_2(\tau)$ and $u_i^\mu(\tau)$ in the form (\ref{uharm})
\be
\begin{array}{c}
\de\s_2(\tau)=\de_0e^{-i\om\tau},\qquad u_2^\mu(\tau)=\big[A_2e^\mu(\tau)+
A_2^\perp\acute e^\mu(\tau)+A_2^ze_3^\mu\big]\,e^{-i\om\tau},\\
u_i^\mu(\tau)=\big[A_i^0e_0^\mu+A_ie^\mu(\tau)-\epsilon_iv_i^{-1}A_i^0
\acute e^\mu(\tau)+A_i^ze_3^\mu\big]\,e^{-i\om\tau},\quad i=1,3.
\rule[3.5mm]{0mm}{1mm} \end{array}
\label{uiA}\ee
Here the conditions (\ref{Uu}) are taken into acount.
This results in the following system of linear equations with respect to
the complex amplitudes $A_i^\nu$, $\de_0$:
$$
\begin{array}{c}
K_1(Q_1S_1+\om C_1)\,A_1^z+K_3(Q_3S_3+\om C_3)\,A_3^z+\om\mu_2A_2^z=0,
\quad K_j(Q_jC_j-\om S_j)\,A_j^z=A_2^z;\rule[3.3mm]{0mm}{1mm}\\
K_1(Q_1\kappa_1C_1-\om S_1)\,A_1^0+i\om Q_1^{-1}C_1A_1=
K_3(Q_3\kappa_3C_3-\om S_3)\,A_3^0-i\om Q_3^{-1}C_3A_3,\rule[3.3mm]{0mm}{1mm}\\
\qquad i\epsilon_jK_j(Q_j\kappa_jS_j+\om C_j)\,A_j^0-\om Q_j^{-1}C_jA_j=
A_2-i\om\de_0,\rule[3.3mm]{0mm}{1mm}\;\;\;j=1,3,\\
i\om A_2+\theta A_2^\perp=0,\quad
\epsilon_jK_j^*(\om S_j-Q_jC_j)\,A_j^0-i(v_j\Gamma_j)^{-1}S_jA_j=
A_2^\perp+\theta\de_0,\rule[3.3mm]{0mm}{1mm}\\
\mu_2(\theta^2-\om^2)\,A_2^\perp+K_1^*\big[(Q_1^2\kappa_1-\om^2)\,C_1
-2\om\theta S_1\big]\,A_1^0=K_3^*\big[(Q_3^2\kappa_3-\om^2)\,C_3
-2\om\theta S_3\big]\,A_3^0.\rule[3.3mm]{0mm}{1mm}
\end{array}
$$
Here $C_j=\cos\om(\s_j-\s_2^{rot})$, $S_j=\epsilon_j\sin\om(\s_j-\s_2^{rot})$,
$Q_j=\Gamma_j\theta v_j$, $K_j=(1-v_j^2)/(\theta v_j)$,
$K_j^*=K_j/(v_j\Gamma_j)$, $\Gamma_j=(1-v_j^2)^{-1/2}$,
$\kappa_j=1+v_j^{-2}$, $\mu_2=m_2K_1/m_1=m_2K_3/m_3$.

Non-trivial solutions of this system exist only if the value $\om$
is a root of the equations
\bea
&\displaystyle
\mu_2\om+\frac{(Q_1Q_3-\om^2)\sin\pi\om+(Q_1+Q_3)\cos\pi\om}
{(Q_1C_1-\om S_1)(Q_3C_3-\om S_3)}=0;&\label{zfreql}\\
&\displaystyle
\mu_2\tilde\om\frac{\tilde\om^2-\theta^2}{\tilde\om^2+\theta^2}=
\frac{(Q_1^2\kappa_1-\tilde\om^2)\,C_1-2\tilde\om Q_1S_1}
{(Q_1^2\kappa_1-\tilde\om^2)\,S_1+2\tilde\om Q_1C_1}+
\frac{(Q_3^2\kappa_3-\tilde\om^2)\,C_3-2\tilde\om Q_3S_3}
{(Q_3^2\kappa_3-\tilde\om^2)\,S_3+2\tilde\om Q_3C_3}.&
\label{pfreql}\eea
They generalize correspondingly Eqs.~(\ref{zfreq}) and (\ref{pfreq})
--- the last equations are limits of Eqs.~(\ref{zfreql}), (\ref{pfreql})
if $\mu_2=m_2=0$.
The roots $\om=\om_n$ of Eq.~(\ref{zfreql}) correspond to oscillations
of the rotating system $q$-$q$-$q$ in $z$- or $e_3$-direction,
and the roots $\tilde\om_n$ of Eq.~(\ref{pfreql}) describe oscillations
in the rotational plane.

From the point of view of the stability problem the most important fact is
the presence (if $m_2\ne0$) of the imaginary root in Eq.~(\ref{pfreql}).
It may be easily found after the
substitution $\om=i\xi$, the corresponding value $\xi=\xi^*\in(0,\pi)$.
The spectrum of an arbitrary quasirotational motion of the $q$-$q$-$q$
configuration has the form similar to Eq.~(\ref{osc}) and contains all
oscillatory modes with frequencies $\om_n$ and $\tilde\om_n$, which are
roots of Eqs.~(\ref{zfreql}), (\ref{pfreql}). If this motion has no certain
symmetry, the exponentially growing mode with the factor
$\exp(-i\om\tau)=\exp(\xi^*\tau)$ is in this spectrum. This proves
the conclusion made in Ref.~\cite{lin} about instability of the rotation
(\ref{rot}) of the system $q$-$q$-$q$ in Lyapunov's sense ---
any small asymmetric perturbation is growing.

\bigskip
\noindent{\bf Conclusion}
\medskip

In the present work the we proved that the classical rotational motions
(\ref{rot}) of the linear string baryon model $q$-$q$-$q$ are unstable.
This analysis doesn't allow to describe the future evolution of this
instability when the amplitudes of growing disturbances are not small.
But this process was studied previously in Ref.~\cite{lin} with using the
suggested numerical methods based upon determination of an arbitrary classical
motion of the $q$-$q$-$q$ system if its initial position in Minkowski space
$R^{1,D-1}$ and initial velocities of string points are given.
These numerical experiments showed the picture of the instability
(any arbitrarily small asymmetric disturbances are growing) and demonstrated
 that the result of its evolution is the complicated motion
with quasi-periodical varying of the distance between
the nearest two quarks. However the minimal value of the mentioned distance
$\Delta R$ does not equal zero, in other words, the system $q$-$q$-$q$ is not
transformed in quark-diquark ($q$-$qq$) one, as was supposed in Ref.~\cite{Ko}.

This picture radically differs from that for the relativistic string with
massive ends. For the latter model\footnote{It is the model of the meson
$q$-$\overline q$ or the baryon in the form $q$-$qq$.} both the numerical experiments in Ref.~\cite{stabPRD} and the analysis
in Sect.~2 of this paper demonstrate the stability of the rotational motion
(\ref{rot}) (in the linear approximation).

For the rotational motions of the ``three-string" model or Y configuration
we also see instability both in the numerical calculations Ref.~\cite{stabPRD}
and in the analytical investigations Ref.~\cite{Exc} of spectra
for quasirotational states. This spectrum contains the branch of oscillatory
states, whose dimensionless frequencies $\tilde\om_n$ are roots of the equation
\cite{Exc}
\be
2\frac{Q_1\tilde\om(\th^2-\tilde\om^2)-i(\tilde\om^2-Q_1^2\kappa_1)(\tilde\om^2+\th^2)}
{(\tilde\om^2-Q_1^2\kappa_1)(\tilde\om^2-\th^2)-4iQ_1\tilde\om(\tilde\om^2+\th^2)}=
\cot\pi\tilde\om.
\label{compfr}\ee
These roots are obligatory complex numbers (except for $\tilde\om=\pm\theta$).
Imaginary parts of them are always positive
so the disturbances of this class (branch) are exponentially growing
in accordance with the factor
$
\exp(-i\tilde\om_n\tau)=\exp(-i\Re\tilde\om_n\tau)\exp(\Im\tilde\om_n\tau).
$
Arbitrary quasirotational motion of the system Y may  also be expanded
in the Fourier series of the type (\ref{rot})  with harmonics of all
classes described above. So only for the disturbances with the special
symmetry (when all amplitudes of the modes (\ref{compfr}) equal zero)
these disturbances do not grow exponentially, in other words  --- the
rotational motion for the three-string configuration
is unstable even in the linear approximation.

The evolution of this instability was calculated numerically in Ref.~\cite{stabPRD}
where we showed that
the picture of motion is qualitatively identical for any
small asymmetric disturbance.
Starting from some point in time the junction begins to move.
The distance between the junction and the rotational center increases
and the lengths of the string segments vary quasiperiodically unless
one of the material points inevitably merges with the junction.

This pictures differs from that for the linear string baryon model
$q$-$q$-$q$. It is, in particular, connected with different
properties of the complex roots of Eq.~ (\ref{pfreql}) and Eq.~ (\ref{compfr}).

Instability of the rotations for the string baryon models $q$-$q$-$q$ and Y
is, of course, their drawback. But it is not ``fatal" drawback, the classic
instability is only one of the features for choosing the most adequate string
baryon model among the existing four ones. The most important consequence
of the rotational instability is in impossibility to quantize the quasirotational
states. This procedure can be developed for the states (\ref{osc}) (the Fourier
series) for the string with massive ends in the stable case. But for the
unstable models $q$-$q$-$q$ and Y this procedure is not permitted.

\smallskip

The work is supported by the Russian Foundation of Basic Research,
grant  00-02-17359.

\end{document}